\documentclass[onecolumn,preprintnumbers,superscriptaddress,nofootinbib,aps,prd,floatfix]{revtex4}

\usepackage{amsmath,amssymb}
\usepackage{dsfont}
\usepackage{graphicx} 
\usepackage{epstopdf} 
\usepackage{slashed}
\usepackage{subfigure}
\usepackage{color}
\usepackage{multirow}
\usepackage{pdflscape}

\usepackage{hyperref}

\usepackage{footmisc}

\usepackage{array}
\newcolumntype{L}[1]{>{\raggedright\let\newline\\\arraybackslash\hspace{0pt}}m{#1}}
\newcolumntype{C}[1]{>{\centering\let\newline\\\arraybackslash\hspace{0pt}}m{#1}}
\newcolumntype{R}[1]{>{\raggedleft\let\newline\\\arraybackslash\hspace{0pt}}m{#1}}

\newcommand{\be}{\begin{eqnarray*}}
\newcommand{\ee}{\end{eqnarray*}}

\newcommand{\bee}{\begin{eqnarray}}
\newcommand{\eee}{\end{eqnarray}}
\newcommand{\beeq}{\begin{equation}}
\newcommand{\eeeq}{\end{equation}}

\newcommand{\tev}{{\text{TeV}}}

\renewcommand{\vec}{\bf}

\newcommand{\flavio}{\texttt{flavio}}

\newcommand{\wilson}{\texttt{wilson}}
\newcommand{\smelli}{\texttt{smelli}}

\newcommand{\Dlr}{\overset\leftrightarrow{D}}
\newcommand{\DlrImu}{\Dlr_\mu \hspace*{-0.16cm}{}^I}

\newcommand{\fref}[1]{Fig.~\ref{fig:#1}} 
\newcommand{\eref}[1]{Eq.~\eqref{eq:#1}}

\newcommand{\sref}[1]{Section~\ref{sec:#1}}
\newcommand{\cref}[1]{Chapter~\ref{ch:#1}}
\newcommand{\tref}[1]{Table~\ref{tab:#1}}

\newcommand{\nnl}{\nonumber \\}

\newcommand{\beq}{\begin{equation}} 
\newcommand{\eeq}{\end{equation}} 
\newcommand{\ba}{\begin{array}}  
\newcommand{\ea}{\end{array}} 
\newcommand{\bea}{\begin{eqnarray}}  
\newcommand{\eea}{\end{eqnarray} }  
\newcommand{\bal}{\begin{align}}
\newcommand{\eal}{\end{align}}   
\newcommand{\bi}{\begin{itemize}}  
\newcommand{\ei}{\end{itemize}}  
\newcommand{\ben}{\begin{enumerate}}  
\newcommand{\een}{\end{enumerate}}  
\newcommand{\bc}{\begin{center}}
\newcommand{\ec}{\end{center}} 
\newcommand{\bt}{\begin{table}}
\newcommand{\et}{\end{table}}  
\newcommand{\btb}{\begin{tabular}}
\newcommand{\etb}{\end{tabular}}

\newcommand{\cO}{{\mathcal O}} 
 
\newcommand{\cL}{{\mathcal L}}

\begin{document}

\title{Flavourful SMEFT likelihood for Higgs and electroweak data}
\begin{abstract} 
We perform an updated fit to LHC Higgs data and LEP electroweak precision tests in the framework of the Standard Model Effective Field Theory (SMEFT). We assume a generic structure of the SMEFT operators without imposing any flavour symmetries. The implementation is released as part of the public global SMEFT likelihood. This allows one to fit parameters of a broad class of new physics models to combined Higgs, electroweak, quark flavour, and lepton flavour observables.
\end{abstract}
%
%
\author{Adam Falkowski} \email{adam.falkowski@th.u-psud.fr}
\affiliation{Universit\'{e} Paris-Saclay, CNRS/IN2P3, IJCLab, 91405 Orsay, France}

\author{David Straub} \email{straub@protonmail.com}
\affiliation{Excellence Cluster Universe, Boltzmannstr. 2, 85748 Garching, Germany}

\maketitle

\section{Introduction}

After the discovery of the 125~GeV boson~\cite{Aad:2012tfa,Chatrchyan:2012xdj}, the Higgs searches at the LHC have turned into precision tests of the Standard Model (SM). 
The SM predicts the coupling strength between each fundamental particle and the Higgs in terms of that particle's mass, without any adjustable parameters.
Beyond the SM (BSM), however, these couplings can be modified, 
therefore measurements of Higgs production cross sections and decay rates may reveal the fundamental theory underlying the SM.   
It is beneficial to describe Higgs coupling modifications in the model-independent language of effective field theory (EFT)~\cite{Carmi:2012yp,Azatov:2012bz,Espinosa:2012ir}. 
Assuming new physics decouples, that is to say the masses of non-SM particles are parametrically larger than the weak scale, the relevant effective theory at energies $E\sim  m_h \approx 125$~GeV is the so-called SMEFT. 
That theory has the same particle content and local symmetry as the SM, but it admits interaction terms (operators) in the Lagrangian with canonical dimensions larger than four.  
The Higgs data can be interpreted as constraints on Wilson coefficients of the  higher-dimensional operators. 

Advantages of the SMEFT are not restricted to LHC Higgs physics. 
The theory offers a universal framework to describe a vast spectrum of precision measurements performed in weak scale colliders such as the LHC, Tevatron, or LEP. 
In particular, it is a perfect language to describe the so-called {\em electroweak precision tests}, that is accurate measurements of $Z$ and $W$ boson properties. 
A straightforward observation is that SMEFT operators may simultaneously affect both Higgs and electroweak observables. 
Therefore it makes sense to combine the two sets in order to increase the constraining power of data~\cite{Corbett:2012ja, Pomarol:2013zra, Dumont:2013wma, Corbett:2013pja, Ellis:2014jta, Falkowski:2015jaa, Butter:2016cvz, Berthier:2016tkq, Ellis:2018gqa, Biekotter:2018rhp, deBlas:2019okz}.
This work represents another step forward in this direction.

In this paper we construct a likelihood for Wilson coefficients of  dimension-6 SMEFT operators using a large set of Higgs and electroweak observables. 
Compared to previous analyses, the main novelties are:
\begin{enumerate}
\item We do not impose any simplifying assumptions about the flavour structure of dimension-6 operators.   
This makes the analysis considerably more difficult, but the advantage is that our results are more general. 
The likelihood we provide can thus be used to constrain a variety of BSM models~\cite{deBlas:2017xtg} beyond the $U(3)^5$ or minimal flavour violation paradigm. 
In particular, it can be applied to models addressing the so-called $b \to s \ell \ell$ anomalies~\cite{Aaij:2019wad,Aaij:2017vbb}, which  must always involve a highly non-trivial flavour structure.  
\item All code employed for our analysis is open source. 
This allows the community to scrutinize, reproduce, and modify our analysis, e.g. when new data becomes available.  
We follow the standard Wilson coefficient exchange format (WCxf) \cite{Aebischer:2017ugx}, such that our results can be easily imported by other analysis codes.   
\item We include the most recent Higgs signal strength combinations from ATLAS and CMS based on up to 80~fb${}^{-1}$ of the Run-2 data~\cite{Sirunyan:2018koj,Aaboud:2018fhh}. 
\end{enumerate}

Simultaneously to interpretations of high-energy collider data, much progress  has been made on the front of  low-energy observables measured at energies well below the weak scale. In Refs.~\cite{Gonzalez-Alonso:2016etj,Falkowski:2017pss} a treasure trove of such data was recast as constraints on the flavour-generic SMEFT.  
Moreover, a public code \smelli{}\footnote{For details, see \url{https://smelli.github.io}.} providing a likelihood function in SMEFT Wilson coefficient space  was released \cite{Aebischer:2018iyb}, based on the \flavio{} observable calculator \cite{Straub:2018kue} and  the \wilson{} tool for running and matching Wilson coefficients \cite{Aebischer:2018bkb}. 
Besides electroweak precision tests, the code includes information from quark flavour physics, lepton flavour physics, low-energy parity violation, and other low-energy precision tests.  
It is desirable to include the Higgs observables into the same framework, as many dimension-6 operators generated by typical BSM models are currently probed {\em only} via LHC Higgs searches. 
As a part of this work, the combined Higgs and electroweak likelihood was incorporated into the global SMEFT  likelihood package \smelli{}, which allows users to  perform combined fits of BSM models to Higgs, electroweak, flavour, and other low-energy data.

This paper is organized as follows. 
In \sref{smeft} we briefly review the SMEFT framework in order to fix our conventions and notation.
In \sref{meth} we summarize the  data used in this analysis and describe our methodology of constructing global likelihoods and parameter fitting. 
In \sref{limits} we give the best fit and confidence intervals for the Wilson coefficients affecting the Higgs and/or electroweak observables at tree level.
\sref{app} contains two simple applications of our flavourful likelihood: one designed to connect and compare to previous work, and the other illustrating the relevance of working with a general flavour structure of dimension-6 operators.

\section{SMEFT framework}
\label{sec:smeft}

We briefly review the SMEFT framework to fix our notation.
We consider the extension of the SM by all independent dimension-6 operators $Q_i$ invariant under the SM gauge symmetries \cite{Buchmuller:1985jz, Grzadkowski:2010es}:
\begin{equation}
  \mathcal L_\text{SMEFT} =
  \mathcal L_\text{SM} +
  \sum_i C_i \,Q_i \, . 
\end{equation}
We work in the Warsaw basis of operators \cite{Grzadkowski:2010es}, selecting the weak basis for fermions where the down-type quark and charged lepton mass matrices are diagonal (coinciding with the \texttt{Warsaw} basis as defined in the WCxf \cite{Aebischer:2017ugx}).
In our conventions the Wilson coefficients $C_i$ have dimensions 
$[\rm mass]^{-2}$. 
Operators with dimensions higher than six, as well as dimension-5 operators (which only generate tiny neutrino masses and have no observable effect on Higgs or electroweak phenomenology) are ignored in this analysis. 
We also ignore CP violating opearators, which affect our observables only at the quadratic level in the corresponding Wilson coefficients.  
The electroweak parameters $g_L$, $g_Y$, $v$ 
are defined in the $\alpha$-$m_Z$-$G_F$ scheme. 
We take into account only tree-level effects of dimension-6 operators, except in the observables where the SM contribution itself appears first at one loop. 
In this approximation, the operators affecting the Higgs
and electroweak precision data are shown in~\tref{ops}. The only four-fermion operator entering is the operator $[Q_{ll}]_{1221}$ that interferes with the SM amplitude in muon decay and thus modifies the relation between the muon lifetime (defining the Fermi constant $G_F$) and the electroweak parameters.\footnote{
The operators $[Q_{ll}]_{ijji}$ and $[Q_{ll}]_{jiij}$ are indistinguishable.
In our conventions, only the ones with $i \leq j$ are included in the Lagrangian. 
Another convention encountered in the literature is that both are included and multiplied by the same Wilson coefficient, in which case our results for $[C_{ll}]_{1221}$ have to be multiplied by $1/2$.}

\begin{table}[tb]
\renewcommand{\arraystretch}{1.6}
\begin{center}
\begin{tabular}{cccccc}
\toprule
$Q_{\varphi D}$ & $\big( \varphi^\dagger D^\mu \varphi \big)^\ast \big( \varphi^\dagger D_\mu \varphi \big)$ 
&
$Q_{\varphi l}^{(1)}$ & $\big( \varphi^\dagger i \Dlr_\mu \varphi \big) \big( \bar l \gamma^\mu l \big)$
&
$Q_{u \varphi}$  & $\big( \varphi^\dagger \varphi \big) \big( \bar q u \widetilde \varphi \big)$
\\
$Q_{\varphi \Box}$ & $\big( \varphi^\dagger \varphi \big) \Box \big( \varphi^\dagger \varphi \big)$
&
$Q_{\varphi l}^{(3)}$ & $\big( \varphi^\dagger i \DlrImu \varphi \big) \big( \bar l \tau^I \gamma^\mu l \big)$
&
$Q_{d \varphi}$  & $\big( \varphi^\dagger \varphi \big) \big( \bar q d \varphi \big)$
\\
$Q_{\varphi B}$ & $\varphi^\dagger \varphi B_{\mu \nu} B^{\mu \nu}$
&
$Q_{\varphi e}$ & $\big( \varphi^\dagger i \Dlr_\mu \varphi \big) \big( \bar e \gamma^\mu e \big)$
&
$Q_{e \varphi}$ & $\big( \varphi^\dagger \varphi \big) \big( \bar l e \varphi \big)$
\\
$Q_{\varphi W}$  & $\varphi^\dagger \varphi W_{\mu \nu}^I W^{I \mu \nu}$
&
$Q_{\varphi q}^{(1)}$ & $\big( \varphi^\dagger i \Dlr_\mu \varphi \big) \big( \bar q \gamma^\mu q \big)$
&
$Q_{l l}$ & $\big( \bar l \gamma_\mu l \big) \big( \bar l \gamma^\mu l \big)$
\\
$Q_{\varphi W B}$  & $\varphi^\dagger \tau^I \varphi W_{\mu \nu}^I B^{\mu \nu}$
&
$Q_{\varphi q}^{(3)}$& $\big( \varphi^\dagger i \DlrImu \varphi \big) \big( \bar q \tau^I \gamma^\mu q \big)$
&
$Q_W$  & $\epsilon^{IJK} W_\mu^{I \nu} W_\nu^{J \rho} W_\rho^{K \mu}$
\\
$Q_{\varphi G}$ & $\varphi^\dagger \varphi G_{\mu \nu}^A G^{A \mu \nu}$
&
$Q_{\varphi u}$ & $\big( \varphi^\dagger i \Dlr_\mu \varphi \big) \big( \bar u \gamma^\mu u \big)$
&
$Q_G$  & $f^{ABC} G_\mu^{A \nu} G_\nu^{B \rho} G_\rho^{C \mu}$
\\
\end{tabular}
\end{center}
\caption{Dimension-6 SMEFT operators in the Warsaw basis  relevant for the Higgs and electroweak observables at tree level.}
\label{tab:ops}
\end{table}

\section{Experimental data and fit methodology}
\label{sec:meth}

We summarize the experimental data used in our analysis. 
\bi
\item {\em The $Z$ and $W$ pole observables.}
These are largely unchanged compared to the analysis of  Ref.~\cite{Efrati:2015eaa}, see Table~1 therein for the original references. 
The only new pieces of experimental information are i) the PDG combination for the  $W$ mass~\cite{Tanabashi:2018oca}, which includes the recent ATLAS  measurement~\cite{Aaboud:2017svj};  ii) the measurements of 
${\Gamma (W \to \tau \nu) \over \Gamma(W \to e\nu)}$
in D0~\cite{Abbott:1999pk}  
and of 
${\Gamma (W\to \mu \nu) \over \Gamma(W\to e\nu)}$
in LHCb~\cite{Aaij:2016qqz};  
iii)  the updated values of $\sigma_{\rm had}^0$ and $\Gamma_Z$ following the recent revision of the LEP-1 integrated luminosity~\cite{Janot:2019oyi}. 
On the other hand, we no longer use the $V_{tb}$ determination from Ref.~\cite{Khachatryan:2014iya} to constrain the $W t b$ vertex, as other dimension-6 operators beyond those in \tref{ops} may affect that measurement.    
\item The total and differential cross sections for $WW$ pair production measured in LEP-2 (Tables~5.3 and 5.6 of Ref.~\cite{Schael:2013ita}). 
These are unchanged compared to the analysis of Ref.~\cite{Falkowski:2014tna}. 
\item Combination of Run-1 ATLAS and CMS measurements of the Higgs signal strength measurements in 21 different channels (Table~8 of Ref.~\cite{Khachatryan:2016vau} with the correlation matrix in Fig.~27, plus the $\mu \mu$ signal strength). 
In addition, we use ATLAS~\cite{Aad:2015gba} and CMS~\cite{Chatrchyan:2013vaa} Run-1 measurements of the signal strength in the $Z \gamma$ channel. 
\item Combined ATLAS measurement of the Higgs signal strength based on the 80~fb${}^{-1}$ of run-2 data (Table~6 of Ref.~\cite{Aad:2019mbh} with the correlation matrix in Fig.~6). We also include the signal strength measurements in the $\mu \mu$~\cite{Aaboud:2017ojs},   
$Z \gamma$~\cite{Aaboud:2017uhw}, and  $c \bar c$~\cite{Aaboud:2018fhh} channels. 
\item Combined CMS measurement of the Higgs signal strength based on the 35.9~fb${}^{-1}$ of run-2 data (Table~3 of Ref.~\cite{Sirunyan:2018koj} with the correlation matrix in the auxiliary material). 
We also include the signal strength measurements in the 
$Z \gamma$~\cite{Sirunyan:2018tbk} and  $c \bar c$ channels~\cite{CMS:2019tbh}.
\ei
In order to take into account the correlations, we have to treat the experimental likelihoods as (multivariate)
Gaussians. 
To this end, we symmetrize asymmetric uncertainties.

SMEFT corrections to electroweak observables are calculated analytically at tree level.
For most of the Higgs observables they are determined numerically using the \texttt{SMEFTsim} model implementation~\cite{Brivio:2017btx} in \texttt{Madgraph}~\cite{Alwall:2014hca}. 
The exceptions are the $gg \to h$ and $h \to \gamma \gamma$ processes where we use analytic formulas, taking into account 1-loop effects due to modified Yukawa and $h V_\mu V_\mu$ couplings.  
Logarithmically enhanced one-loop corrections proportional to the dimension-6 Wilson coefficient are included in our code via the renormalization group running effects~\cite{Alonso:2013hga}.
Finite loop corrections (see e.g.~\cite{Dawson:2019clf,Dawson:2018liq,Dedes:2018seb,Dawson:2018pyl,Hartmann:2015aia}) other than those described above are ignored in our analysis. 
Our calculation of the Higgs and diboson observables include the effects due to  modified $Z$ and $W$ couplings to fermions~\cite{Zhang:2016zsp,Banerjee:2018bio}. 
We take into account the interference effects in 
$h \to V^{(*)}V^* \to 4 f$ decays~\cite{Chen:2013ejz,Falkowski:2015fla,Brivio:2019myy}. 

Having expressed all observables as linear functions
of SMEFT Wilson coefficients, $\vec{O}_\text{th}(C_i)$,  we construct the likelihood function $L(C_i)=e^{-\chi^2(C_i)/2}$, with
\begin{equation}
  \chi^2(C_i) = \vec{x}(C_i)^T \,S^{-1} \, \vec{x}(C_i),
\label{eq:chi2}
\end{equation}
where $S$ is the experimental covariance matrix
and
\begin{equation}
  \vec{x}(C_i)=\vec{O}_\text{th}(C_i)-\vec{O}_\text{exp}
\end{equation}
is the difference between the observables predicted in the SMEFT and measured experimentally.
Given the precision of SM calculations of the observables
in question, we can neglect all theory uncertainties and
thus obtain a likelihood function that only depends on
Wilson coefficients, and not on nuisance parameters.

\section{Confidence intervals for Wilson coefficients}
\label{sec:limits}

In this section we present marginalized 1~$\sigma$ confidence intervals for the Wilson coefficients entering into our combined likelihood.
This exercise illustrates, in a model-independent fashion,  the constraining power of the current Higgs and electroweak data.  

The electroweak and Higgs constraints dramatically differ in accuracy:  for the former the typical precision is $\cO(0.1\%)$, while for the latter it is $\cO(10\%)$. 
In the Warsaw basis the two sets of constraints probe an overlapping set of Wilson coefficients, which leads to large correlations.  
For the sake of illustration, it is more transparent to work with  certain linear combinations of Wilson coefficients, such that the strongly constrained combinations are isolated from the weakly constrained ones~\cite{Pomarol:2013zra,Gupta:2014rxa}.  
With this in mind, we define\footnote{%
Note that our $\delta g$'s differ by $1/v^2$ from the ones defined in Refs.~\cite{Efrati:2015eaa,Falkowski:2017pss}.}
the following linear combinations of the Warsaw basis Wilson coefficients: 
\bea
\label{eq:CI_deltagzf}
\delta g^{W \ell}_L & = &   C^{(3)}_{\varphi l} + f(1/2,0) - f(-1/2,-1),  
\nnl 
\delta g^{Z\ell}_L & = &    - {1 \over 2} C^{(3)}_{\varphi l} - {1\over 2} C_{\varphi l}^{(1)}+   f(-1/2, -1) , 
\nnl 
\delta g^{Z\ell }_R & = &  - {1\over 2} C_{\varphi e}^{(1)}   +  f(0, -1) ,
\nnl 
\delta g^{Z u}_L & = &   {1 \over 2}  C^{(3)}_{\varphi q} - {1\over 2} C_{\varphi q}^{(1)}   + f(1/2,2/3) ,  
\nnl
\delta g^{Zd}_L & = &   
 - {1 \over 2}  C^{(3)}_{\varphi q} - {1\over 2} C_{\varphi q}^{(1)}   + f(-1/2,-1/3), 
\nnl
\delta g^{Zu}_R & = &    - {1\over 2} C_{\varphi u}   +  f(0,2/3), 
\nnl
\delta g^{Zd}_R & = &    - {1\over 2} C_{\varphi d}  +  f(0,-1/3), 
\eea 
where 
\beq 
\label{eq:CI_fdeltag}
f(T^3,Q)  \equiv  \bigg \{  
-   Q  {g_L  g_Y \over  g_L^2 -  g_Y^2} C_{\varphi WB} 
 -  {\bf 1}  \left ( 
{1 \over 4} C_{\varphi D}  +  {1 \over 2 } \Delta_{G_F}  \right )  \left ( T^3 + Q { g_Y^2 \over  g_L^2 -   g_Y^2} \right )  \bigg \} {\bf 1},
\eeq
and 
$\Delta_{G_F} = [C^{(3)}_{Hl}]_{11} + [C^{(3)}_{Hl}]_{22} -  {1 \over 2}[C_{ll}]_{1221}$. 
In fact, $v^2 \delta g^{Vf}$ are the vertex corrections parametrizing deviations from the SM prediction of the $V=Z,W$ boson couplings to fermion $f$.  
Since they are probed by the LEP electroweak precision tests,  almost all $\delta g$'s are independently constrained with a good accuracy, except for one weakly constrained direction in the space of the light quark vertex corrections~\cite{Efrati:2015eaa}. 

We also define another set of linear combinations of the Warsaw basis Wilson coefficients
\bea
\label{eq:CI_cv}
\delta c_{z} & = & 
C_{\varphi\Box}   - {1 \over 4} C_{\varphi D}
 -{3 \over 2}  \Delta_{G_F}, 
 \nnl 
c_{z\Box} &= &    {1 \over 2   g_L^2} \left (
C_{\varphi D}   
 +2  \Delta_{G_F}   \right ),   
\nnl 
c_{gg} &= & {4 \over  g_s^2} C_{\varphi G},  
\nnl 
c_{\gamma \gamma} &= &  4  \left ( {1  \over   g_L^2} C_{\varphi W} + { 1 \over   g_Y^2} C_{\varphi B} - {1 \over   g_L   g_Y}  C_{\varphi WB} \right )  , 
\nnl  
c_{zz} &= &    4 \left (  {  g_L^2 C_{\varphi W} +    g_Y^2 C_{\varphi B} +     g_L   g_Y  C_{\varphi WB} \over (  g_L^2 +   g_Y^2)^2} \right ),
\nnl 
c_{z\gamma} &= &     4 \left ( { C_{\varphi W} -   C_{\varphi B} 
-  {  g_L^2 -    g_Y^2 \over 2   g_L   g_Y} C_{\varphi WB} \over   g_L^2 +   g_Y^2} \right )  . 
\eea 
The important point is that, unlike $\delta g$, the combinations  $c_i$ defined above  do {\em not} affect electroweak observables at tree level. 
Therefore they are only loosely constrained by the less precise Higgs data, and they are weakly correlated with $\delta g$'s.  

We now present the best fit for the Wilson coefficients after a couple of simplifying but physically reasonable assumptions. 
First, we ignore the dependence of the likelihood on the Wilson coefficients affecting only the Yukawa couplings of the light fermions: $[C_{u\varphi}]_{ii}$, $[C_{d\varphi}]_{ii}$, $i=1,2$, and $[C_{e\varphi}]_{11}$. Currently, these are probed mostly via their contributions to the total Higgs width (thus affecting all observed branching fractions uniformly), and their effect is negligible unless the shift of the Yukawa couplings exceeds the SM Yukawa by orders of magnitude. 
Leaving these parameters in the fit would lead to flat directions.  
We also ignore the contributions to  Higgs observables  proportional to the Wilson coefficient $c_G$, which enters into our likelihood via its contributions to the $t \bar t h$ production. 
This parameter is much better constrained by dijet observables~\cite{Krauss:2016ely},  and after taking that  into account it cannot affect the $t \bar t h$ signal strength significantly.   
With the above assumptions,  {\em 31 independent combinations of Wilson coefficients} are left as free parameters in the fit.  
Our results are shown in \tref{fit}.
As advertised, the combinations $\delta g$ are more strongly constrained, at the level $|\delta g| \lesssim 10^{-2}$--$10^{-3}$, compared to the other combinations probed only by Higgs observables.  
There are a few exceptions from this rule, however.
First, the uncertainty for the combinations $\delta g^{Zq}$ corresponding to light quark vertex corrections are larger~\cite{Efrati:2015eaa} . 
While LEP-1 measures the total hadronic width with a per mille precision, it does not resolve all light quark couplings to $Z$ independently, leading to an approximate flat direction that is only lifted by less precise measurements. 
Second, some of the Wilson coefficients affecting Higgs observables only are strongly constrained when they compete with the SM loop-induced processes ($c_{gg}$, $c_{\gamma \gamma}$), or with small SM Yukawa couplings ($[C_{e\varphi}]_{22}$, $[C_{e\varphi}]_{33}$, $[C_{d\varphi}]_{33}$).

\begin{table}[tb]
\renewcommand{\arraystretch}{1.2}
\centering
\begin{tabular}{lrclc}
\hline
Coeff. & central && unc. & pull [$\sigma$]\\
\hline
$c_W                       $ & $0.053   $ & $\pm$ & $1.8   $ &
$0.0$ \\
$c_{\gamma\gamma}          $ & $-0.048  $ & $\pm$ & $0.1   $ & $0.5$ \\
$c_{gg}                    $ & $-0.017  $ & $\pm$ & $0.014 $ & $1.2$ \\
$c_{z\Box}                 $ & $4.4     $ & $\pm$ & $6.0   $ & $0.7$ \\
$c_{z\gamma}               $ & $-0.58   $ & $\pm$ & $0.61  $ & $0.9$ \\
$c_{zz}                    $ & $-10.0   $ & $\pm$ & $15.0  $ & $0.7$ \\
$\delta c_z                $ & $0.44    $ & $\pm$ & $1.5   $ & $0.3$ \\
$\delta g^{Wl}_{L\,11}     $ & $-0.083  $ & $\pm$ & $0.052 $ & $1.6$ \\
$\delta g^{Wl}_{L\,22}     $ & $-0.23   $ & $\pm$ & $0.082 $ & $2.8$ \\
$\delta g^{Wl}_{L\,33}     $ & $0.27    $ & $\pm$ & $0.099 $ & $2.8$ \\
$\delta g^{Ze}_{L\,11}     $ & $-0.0056 $ & $\pm$ & $0.0052$ & $1.1$ \\
$\delta g^{Ze}_{L\,22}     $ & $0.00043 $ & $\pm$ & $0.018 $ & $0.0$ \\
$\delta g^{Ze}_{L\,33}     $ & $-0.0036 $ & $\pm$ & $0.01  $ & $0.4$ \\
$\delta g^{Zd}_{L\,11}     $ & $-0.072  $ & $\pm$ & $0.61  $ & $0.1$ \\
$\delta g^{Zd}_{L\,22}     $ & $0.34    $ & $\pm$ & $0.53  $ & $0.6$ \\
$\delta g^{Zd}_{L\,33}     $ & $0.054   $ & $\pm$ & $0.027 $ & $2.0$ \\
$\delta g^{Zd}_{R\,11}     $ & $1.4     $ & $\pm$ & $0.94  $ & $1.5$ \\
$\delta g^{Zd}_{R\,22}     $ & $0.29    $ & $\pm$ & $0.72  $ & $0.4$ \\
$\delta g^{Zd}_{R\,33}     $ & $0.33    $ & $\pm$ & $0.11  $ & $3.1$ \\
$\delta g^{Ze}_{R\,11}     $ & $-0.0097 $ & $\pm$ & $0.0055$ & $1.8$ \\
$\delta g^{Ze}_{R\,22}     $ & $-0.0025 $ & $\pm$ & $0.021 $ & $0.1$ \\
$\delta g^{Ze}_{R\,33}     $ & $0.0074  $ & $\pm$ & $0.01  $ & $0.7$ \\
$\delta g^{Zu}_{L\,11}     $ & $0.17    $ & $\pm$ & $0.47  $ & $0.4$ \\
$\delta g^{Zu}_{L\,22}     $ & $-0.032  $ & $\pm$ & $0.072 $ & $0.5$ \\
$\delta g^{Zu}_{R\,11}     $ & $1.2     $ & $\pm$ & $0.78  $ & $1.6$ \\
$\delta g^{Zu}_{R\,22}     $ & $-0.055  $ & $\pm$ & $0.084 $ & $0.7$ \\
$[C_{u\varphi}]_{33}       $ & $-0.28   $ & $\pm$ & $2.7   $ & $0.1$ \\
$[C_{d\varphi}]_{33}       $ & $0.018   $ & $\pm$ & $0.049 $ & $0.4$ \\
$[C_{e\varphi}]_{22}       $ & $0.004   $ & $\pm$ & $0.0053$ & $0.8$ \\
$[C_{e\varphi}]_{33}       $ & $-0.00059$ & $\pm$ & $0.018 $ & $0.0$ \\
$[C_{ll}]_{1221}           $ & $-0.64   $ & $\pm$ & $0.2   $ & $3.2$ \\
\hline
\end{tabular}
\caption{Best-fit values (in units of TeV$^{-2}$) and pulls
of the Warsaw basis Wilson coefficient and their linear  combinations defined in \eref{CI_deltagzf} and \eref{CI_cv}.
\label{tab:fit}
}
\end{table}

The best fit point has $\Delta \chi^2 = \chi^2_{\rm SM} - \chi^2_{\rm min} \approx 36.7$, which translates to a p-value of approximately $20\%$ for the SM hypothesis.  
There is no significant hint of physics beyond the SM in the
fit, even though some of the individual couplings
in \tref{fit} display pulls (defined simply as the number of
Gaussian standard deviations away from 0) of order $3\sigma$.

The correlation matrix is shown in \fref{corr}. 
The change  of variables in \eref{CI_deltagzf} and \eref{CI_cv} disentangles most of the large correlations present if the original Warsaw basis variables. 
Some $\cO(1)$ correlations remain, however. 
Notably, there are large correlations between the combinations $c_{zz}$ and $c_{z \Box}$. 
This is due to the fact that the current Higgs data poorly disentangle different possible Lorentz structures of the Higgs couplings to electroweak gauge bosons. 
This situation can be somewhat improved by including in the analysis, in addition to the signal strength,
transverse-momentum distributions in the gluon fusion or invariant mass distributions in the associated Higgs production~\cite{Corbett:2015ksa,Englert:2015hrx,Grazzini:2016paz,Banerjee:2018bio}. 
We note that, for strongly correlated Wilson coefficients, the best fit values and the magnitude of the errors may be sensitive to including in the observables quadratic ($\cO(\Lambda^{-4})$) effects in Wilson coefficients. Indeed, for $c_{zz}$ and $c_{z \Box}$ we find that the error change by $\sim 50\%$ upon including the quadratic corrections, with smaller or negligible effects for other Wilson coefficients.   

\begin{figure}[tb]
  \includegraphics[width=0.85 \textwidth]{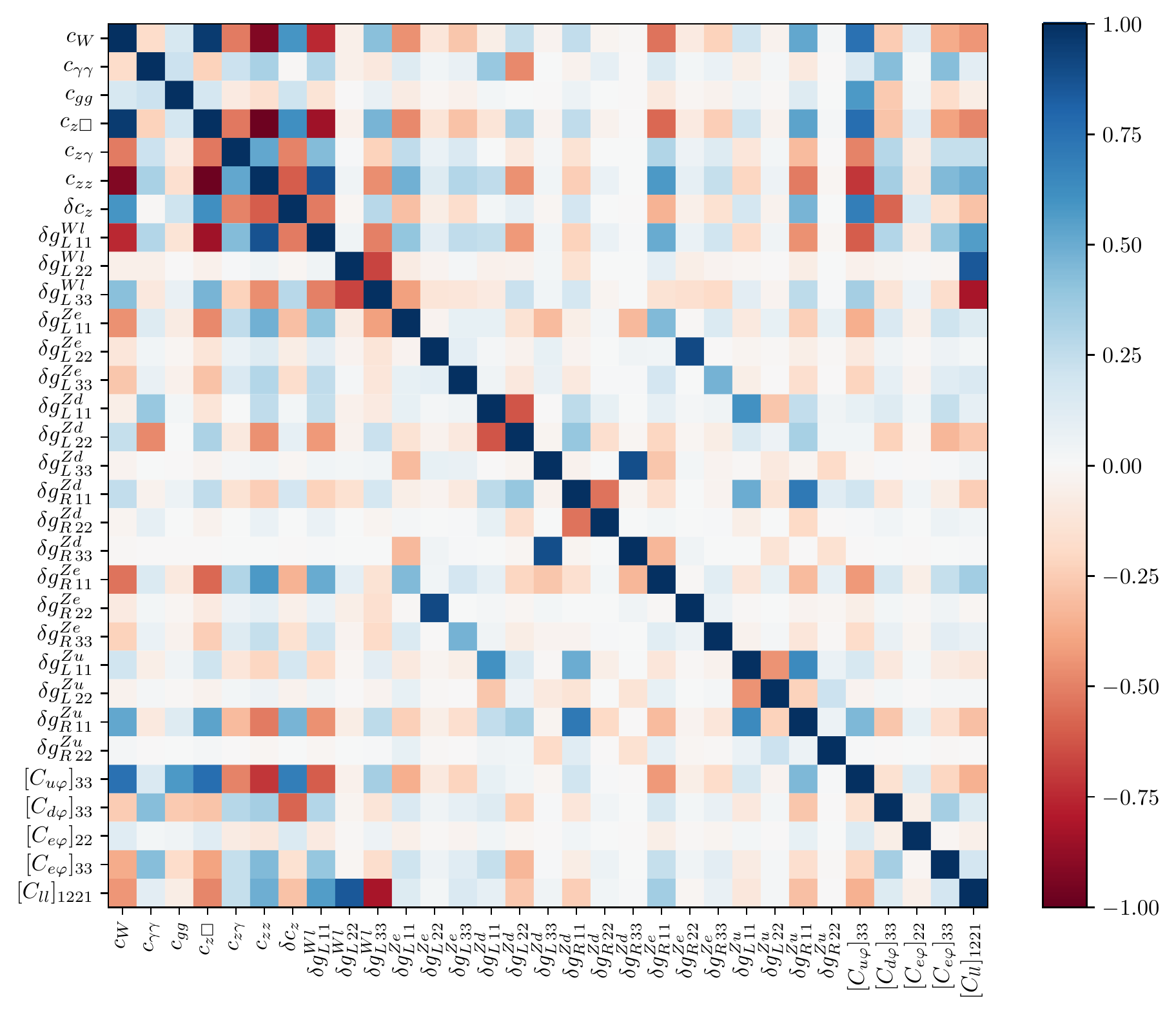}
  \caption{The correlation matrix for the constraints on Wilson coefficients in  the 31-parameter fit in \tref{fit}.}
  \label{fig:corr}
\end{figure}

In closing we remark that it is possible to further relax the assumptions of this global fit without losing a stable minimum. Namely, it is possible to leave also the Wilson coefficient $[C_{u\varphi}]_{22}$ as a free parameter in the fit.
This is thanks to the direct measurement of the $h\to c \bar c$ signal strength at the LHC, which constrains the possible magnitude of charm Yukawa coupling modifications due to $[C_{u\varphi}]_{22}$. 
In the relaxed 32-parameter fit the results are mostly unchanged with respect to those displayed in \tref{fit}, except for a threefold increase of the error on the parameter $\delta c_{z}$. That increase happens because introducing $[C_{u\varphi}]_{22}$ opens an approximately flat direction corresponding to simultaneously increasing both the total Higgs width and the $h V_\mu V_\mu$ couplings. 
The flat direction is lifted precisely thanks to the direct $h\to c \bar c$ measurement~\cite{Aaboud:2018fhh,CMS:2019tbh}.

\section{Applications}
\label{sec:app}

\subsection{Oblique parameters}

Our first example application is the fit to oblique parameters starting from our general likelihood. 
To this end we assume that, at the energy scale $\mu \sim m_Z$, only two Wilson coefficients $C_{\varphi WB}$ and $C_{\varphi D}$ are non-negligible. 
We set all the remaining parameters to zero, and minimize the resulting two-dimensional likelihood.  
This procedure is directly related to the classic assumption that new physics enters via the so-called oblique $S$ and $T$ parameters~\cite{Peskin:1991sw,Wells:2015uba}, with the identification  
\beq
\label{eq:stmap}
C_{\varphi WB} = {g_L g_Y \over 16 \pi v^2} S,    \qquad 
C_{\varphi D}  = - {g_L^2 g_Y^2 \over 2 \pi (g_L^2 + g_Y^2) v^2} T . 
\eeq 
Obviously, this simple example is not using the full flavourful power of our approach. 
Nevertheless it is useful to present here in order to connect and compare to previous works.  
For the Wilson coefficients at the scale $m_Z$ we find 
\beq
C_{\varphi WB}  =  0.0027 \pm 0.0028~{\rm TeV}^{-2},  
\qquad C_{\varphi D} = -0.0170 \pm 0.0094~{\rm TeV}^{-2} ,
\eeq 
with the correlation coefficient $\rho = -0.74$. 
This translates to $S= 0.035 \pm 0.038$, $T= 0.066 \pm 0.036$. 
The best fit ellipses are shown in \fref{st}, for the combined likelihood, and for the electroweak and Higgs likelihoods separately. It can be seen that the LHC Higgs data contribute to constraining the $S$ parameter, mostly via measurements of the $h \to \gamma \gamma$ rate~\cite{Ellis:2018gqa}. 

\begin{figure}[tb]
  \includegraphics[width=0.50 \textwidth]{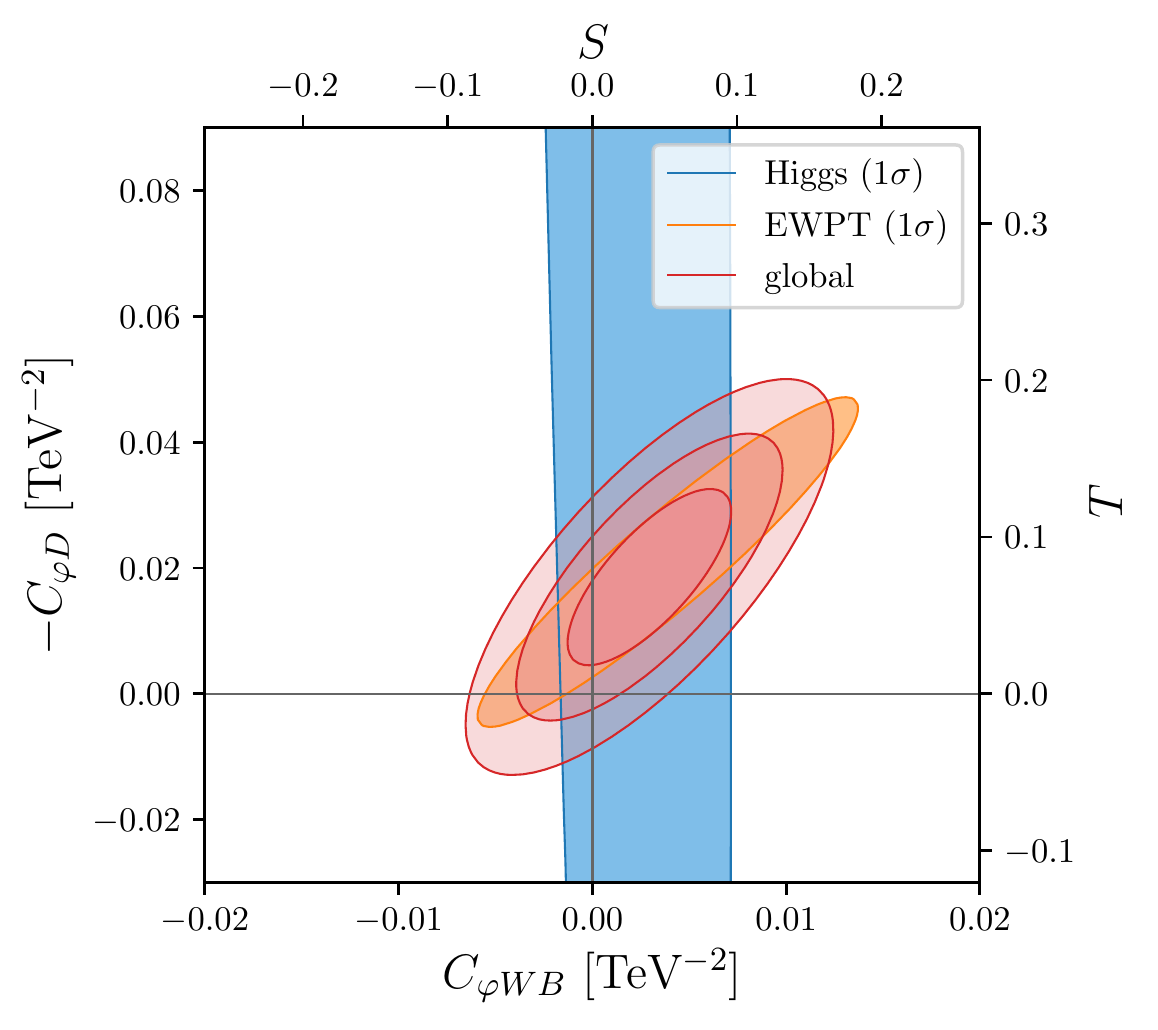}
  \caption{ {\bf S-T fit} using our combined Higgs and electroweak likelihood.
We assume the only non-negligible Wilson coefficients of dimension-6 operators at the scale $m_Z$ are $c_{\varphi D}$ and $c_{\varphi WB}$. 
This is equivalent to the usual fit to the oblique $S$ and $T$ parameters via the map in \eref{stmap}. 
We show the $1\sigma$ preferred region separately for Higgs (blue) and electroweak (orange) data. 
The red contours mark the $1\sigma$, $2\sigma$, and $3\sigma$ preferred regions using the combined likelihood.  
  }
  \label{fig:st}
\end{figure}

\subsection{Custodial vector resonance model}

Another example we consider is the model with an $SU(2)$ triplet $V_\mu^I$ of  massive vector resonances coupled to the SM Higgs, lepton $l$ and quark $q$ doublets as 
\beq
\label{eq:model}
\cL \supset {1 \over 2} V_\mu^I \left ( 
i g_H H^\dagger \tau^I D_\mu H
- i g_H D_\mu H^\dagger \tau^I H
+ \sum_{i} g_{l_i} \bar l_i \tau^I \gamma^\mu  l_i 
+ \sum_{i} g_{q_i} \bar q_i \tau^I \gamma^\mu q_i 
\right ), 
\eeq 
where $i=1,2,3$ is the SM generation index, and we allow the couplings to be flavour-non-universal. 
This kind of resonances and interactions arises e.g. in composite Higgs or warped extra-dimensional scenarios.
The parameter space of our simplified model is characterized by 7 couplings $g_k$ and the resonance mass $M$.  
Assuming $U(3)_q \times U(3)_l$ flavour symmetry would reduce the number of independent couplings to three: $g_H$, $g_l$ and $g_q$. 
Integrating out the massive resonance leads to the SMEFT with the following Wilson coefficients of the operators in \tref{ops}: 
\beq
[C_{\varphi l}^{(3)}]_{ii}  = - {g_H g_{l_i} \over 4 M^2}, \ \
[C_{\varphi q}^{(3)}]_{ii}  = - {g_H g_{q_i} \over 4 M^2}, \ \
C_{\varphi \Box } = -3 {g_H^2 \over 8 M^2},\ \ 
[C_{f \varphi}]_{ii} = - {g_H^2 y_{f_i} \over 4 M^2}, \ \ 
[C_{ll}]_{1221} =  -{g_{l_1} g_{l_2} \over 2 M^2}. 
\eeq 
where $y_{f_i} = {\sqrt 2 m_{f_i} \over v}$ is the Yukawa coupling of the fermion $f_i$, $f = u,d,l$.
As usual, only the ratios coupling/mass are available to a low-energy observer. 
Thus the SMEFT parameter space describing our simplified model is 7-dimensional in the generic case, and 3-dimensional in the  $U(3)_q \times U(3)_l$ limit.  
We ignore the effects of the operator $Q_{\varphi}$, which only affects double Higgs production and  is very weakly constrained at present. 
Four-fermion operators other than $[Q_{ll}]_{1221}$, also generated in this model, are neglected in this analysis. 
They do not enter into the Higgs and electroweak observables at tree level, but they may affect other precision observables (e.g. LEP-2 fermion scattering, low-energy parity violation, Drell-Yan production at the LHC). 
Ignoring these effects is justified assuming $|g_H| \gg |g_{f_i}|$. 
Such coupling hierarchies can arise naturally in the composite/extra-dimensional scenario, e.g. via fermion localization in an extra dimension. 
We note that, in the limit $|g_H| \gg |g_{f_i}|$,  
constraints from the Higgs observables can be comparable or superior to those from the electroweak observables, which further motivates combining the two.

In this simplified model, one observes a profound difference between the allowed parameter space in the flavor-symmetric and generic cases. 
In the  $U(3)_q \times U(3)_l$ case we find the $1\sigma$ confidence intervals 
\beq
{g_H \over M} = 0.236^{+0.086}_{-0.098}~{\tev}^{-1}, 
\qquad 
{g_l \over g_H} = 0.057 \pm 0.039, 
\qquad 
{g_q \over g_H} = 0.014 \pm 0.040  .
\eeq 
On the other hand, for generic flavour-dependent couplings the limits can be weaker: 
\beq
{g_H \over M} =  0.231^{+0.067}_{-0.107}~{\tev}^{-1}, 
\qquad 
{g_{l_1} \over g_H} = 0.054 \pm 0.040, 
\quad 
{g_{l_2} \over g_H} = 0.100 \pm 0.060, 
\quad 
{g_{l_3} \over g_H} = 0.009 \pm 0.068, 
\nonumber 
\eeq 
\beq
{g_{q_1} \over g_H} = -0.05 \pm 0.25, 
\quad 
{g_{q_2} \over g_H} = 0.11 \pm  0.27, 
\quad
{g_{q_3} \over g_H} = -0.027 \pm 0.094. 
\eeq 
We see the for the resonance coupling to the Higgs field the limits are fairly independent on how the resonance couples to the SM fermions. 
This is because the limit is dominated  by the Higgs data.  
On the other hand, the limits on the resonance couplings to the SM fermions are sensitive to whether or not $U(3)_q \times U(3)_l$ is assumed, 
as also illustrated in \fref{vector}. 
The difference is most dramatic for the couplings to the first two generation of quarks, which are allowed to be an order of magnitude larger in the generic case than in the $U(3)_q \times U(3)_l$ case. 
\begin{figure}[tb]
  \includegraphics[width=0.7 \textwidth]{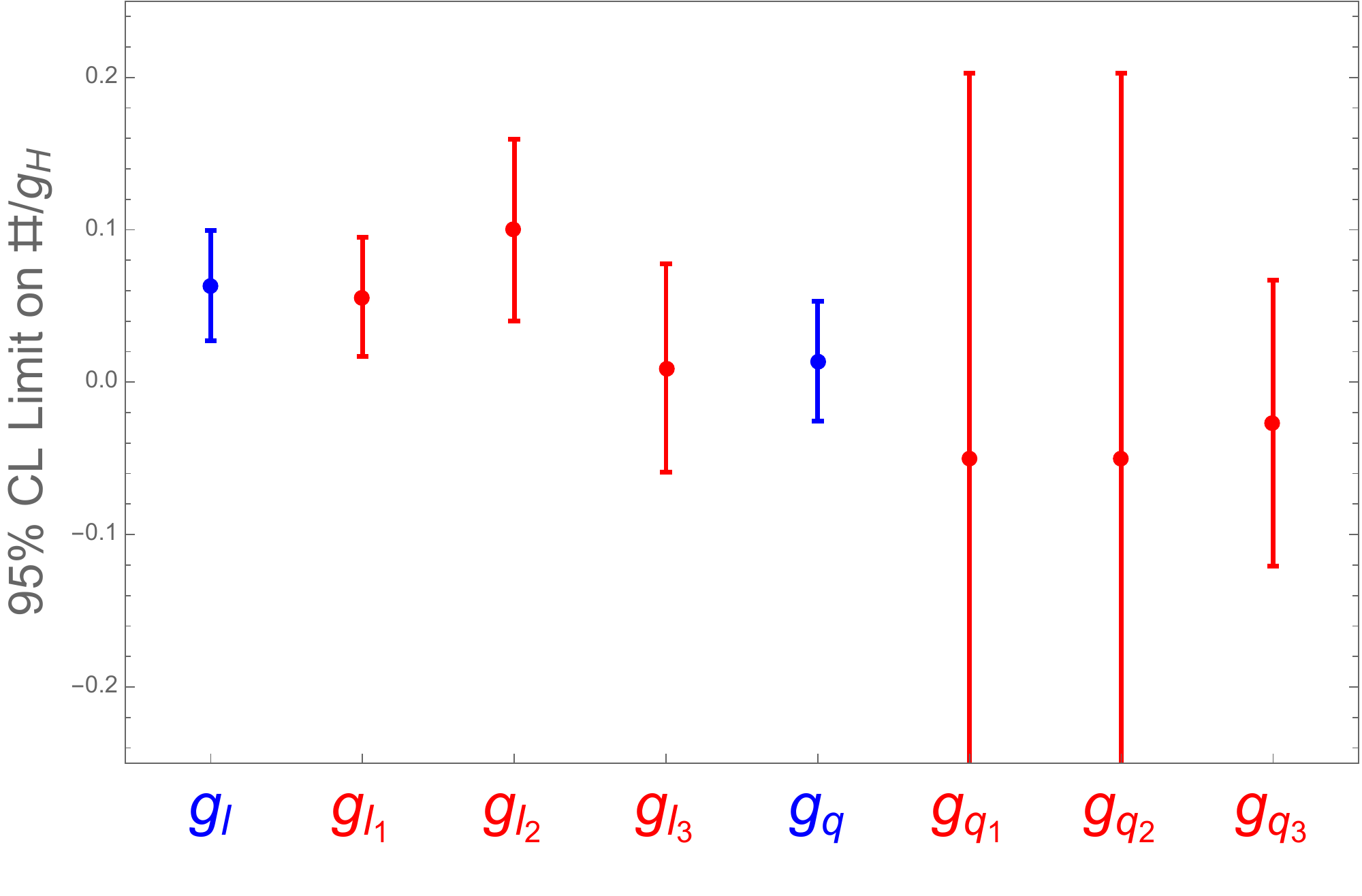}
  \caption{  
Comparison of the limits on the parameter space of the $SU(2)$ vector resonance model in the  $U(3)_q \times U(3)_l$ flavour universal (blue) and flavour generic (red) cases. 
We display the 95\% CL confidence intervals for the resonance couplings to the SM fermions $g_{f_i}$ normalized to its coupling to the Higgs $g_H$, see \eref{model}. 
In the flavour universal case we have $g_{f_i} = g_f$ for all three generations, in which case the bounds are much stronger than in the flavour generic case when $g_{f_1} \dots g_{f_3}$ are all independent. 
For the coupling $g_H$, the bounds are dominated by the LHC Higgs data and thus weakly depend on the flavour assumptions. 
We find the 95\% CL limits: 
$|g_H| \leq 0.97~\tev^{-1}$ in the flavour-universal case and  $|g_H| \leq 0.82~\tev^{-1}$ in the case of generic flavor-dependent couplings. 
 }
  \label{fig:vector}
\end{figure}

\section{Summary}
\label{sec:conc}

In this paper we presented an updated fit of the dimension-6 SMEFT operators to combined Higgs and electroweak precision data. We included the most recent ATLAS and CMS combinations of run-2 Higgs data, and also complemented the LEP electroweak data with a couple of recent measurements in hadronic machines.   
The analysis is based on an open source Python code, which allows the users to easily to scrutinize, reproduce, or update our analysis. The code and conventions follow the WCxf format, so that our results can easily be imported by other analysis programs. 

At the physics level, the main improvement compared to earlier works is that we allow for a completely general flavour structure of dimension-6 operators. 
Thanks to that, the likelihood we provide can be used to constrain a broad class of BSM models beyond the $U(3)^5$ or minimal flavour violation paradigm. 

\acknowledgements

AF is partially supported by the European Union's Horizon 2020 research and innovation programme under the Marie Sk\l{}odowska-Curie grant agreements No 690575 and No 674896, and by the French Agence Nationale de la Recherche (ANR) under grant ANR-19-CE31-0012 (project MORA). 
We thank Ken Mimasu,  Michael Spannowsky, and Michael Trott for useful discussions.
This research was supported by the Munich Institute for Astro- and Particle Physics (MIAPP) of the DFG Excellence Cluster Origins (www.origins-cluster.de). 

\appendix

\bibliographystyle{./JHEP}
\bibliography{main}

\end{document}